\documentclass[aps,prb,showpacs,floatfix,superscriptaddress,raggedbottom,
nobalancelastpage,amssymb,twocolumn]{revtex4-1}
\usepackage{amsmath}
\usepackage{amsfonts}
\usepackage{graphicx}
\usepackage{appendix}
\usepackage{dsfont}
\usepackage{amssymb}
\usepackage{bm}
\usepackage{color}
\usepackage{soul}
\usepackage{natbib}
\usepackage{nccmath}
\usepackage{array}
\usepackage{enumitem}

\usepackage{multirow}
\usepackage{wasysym}


\newcommand{\be}{\begin{equation}}
\newcommand{\ee}{\end{equation}}
\newcommand{\bes}{\begin{equation*}}
\newcommand{\ees}{\end{equation*}}
\newcommand{\bea}{\begin{eqnarray}}
\newcommand{\eea}{\end{eqnarray}}
\newcommand{\ba}{\begin{eqnarray*}}
\newcommand{\ea}{\end{eqnarray*}}

\newcommand{\dis}{\displaystyle}

\newcommand{\fract}[2]{\frac{\dis #1}{\dis #2}}
\newcommand{\Tr}{\mathrm{Tr}}
\newcommand{\eqn}[1]{(\ref{#1})}
\newcommand{\ket}[1]{\mid\! #1\rangle}
\newcommand{\bra}[1]{\langle #1\!\mid}

\newcommand{\ep}{{\epsilon}}

\newcommand{\bw}{\begin{widetext}}
\newcommand{\ew}{\end{widetext}}
\newcommand{\esp}[1]{\text{e}^{#1}}
\newenvironment{eqso}%
{\begin{equation} \begin{aligned}}%
{\end{aligned} \end{equation} }
\newcommand{\beal}{\begin{eqso}}
\newcommand{\eal}{\end{eqso}}
\newenvironment{eqss}%
{\begin{equation*} \begin{aligned}}%
{\end{aligned} \end{equation*} }
\newcommand{\beals}{\begin{eqss}}
\newcommand{\eals}{\end{eqss}}
\newcommand{\bd}[1]{\boldsymbol{#1}}
\newcommand{\mb}{\mathbf{m}}
\newcommand{\bsigma}{\boldsymbol{\sigma}}

\begin{document}
\title{Lindblad dissipative dynamics in presence of phase coexistence} 
\author{Andrea Nava}  
\affiliation{International School for Advanced Studies (SISSA), Via Bonomea 265, I-34136 Trieste, Italy} 
\author{Michele Fabrizio}
\affiliation{International School for Advanced Studies (SISSA), Via Bonomea 265, I-34136 Trieste, Italy} 
\date{\today}
\begin{abstract}
We investigate the dissipative dynamics yielded by the Lindblad equation 
within the coexistence region around a first order phase transition. 
In particular, we consider an exactly-solvable fully-connected 
quantum Ising model with $n$-spin exchange ($n>2$) -- the prototype of quantum 
first order phase transitions -- and several variants of the Lindblad equations. We show that physically sound results, including exotic non-equilibrium phenomena like the Mpemba effect, can be obtained only when 
the Lindblad equation involves jump operators defined for each of the coexisting phases, whether stable or metastable.  
 
  
\end{abstract}
\pacs{    } 
\maketitle

\section{INTRODUCTION}

Even if it covers 71\% of the earth surface and more than half of human body 
is made of it, water still represents a scientific challenge 
for its fascinating and intriguing physical properties \citep{leidenfrost,renfrew}, among which its  
anomalous relaxation time, revealed, e.g., by  
hot buckets of water freezing faster than colder ones, both 
exposed to the same subzero environment. This counter intuitive effect,
already known by Aristotle, Bacon and Descartes, was formalized only
in 1969 by the high school student Erasto Batholomeo Mpemba in the
attempt to freeze a hot ice cream mixture\cite{mpemba}, after him named Mpemba effect. Although
the existence of such phenomenon in water has been questioned\cite{burridge},
it has actually been observed also in clathrate hydrates\cite{Ahn2016}, magneto-resistance
alloys\cite{alloys}, granular systems\cite{PhysRevLett.119.148001}
and spin glasses\cite{spinglasses}. Several proposals have been
formulated to explain how and when the Mpemba effect may or not take
place, which invoke properties of the hydrogen bonds
\citep{zhang}, evaporation\cite{Firth_1971,doi:10.1119/1.1975687},
conduction and convection\cite{VYNNYCKY2015243}, and, eventually, 
the supercooling process\cite{doi:10.1119/1.18059,ESPOSITO2008757}.\\
Usually, liquid water cooled down to the freezing temperature 273.15 K starts
to crystallize around nucleation sites. As discovered
by D. G. Fahrenheit in 1724\cite{fahrenheit}, in pure, i.e, free of impurities, water 
it is possible to delay by a proper cooling procedure the formation of ice nucleators, and thus 
the freezing, till temperatures of the
order of 231.15 K, the liquid spinodal temperature. 
Supercooled water remains trapped in the liquid metastable state 
for quite a long time before spontaneously crystallising in the stable solid phase, unless 
shaken. When supercooling is realized during
a fast process, like a quench, the Newton\textquoteright s
heat law may not apply and the Mpemba effect occur. Specifically, in the
free-energy landscape corresponding to the solid-liquid coexistence region, a system cooled down 
from a lower temperature could fall in the metastable minimum with higher probability and 
remain trapped there for a longer time than a system cooled down from a higher temperature. \\

At low temperature, a faithful description of supercooling, and its associated phenomena like, e.g., the Mpemba effect, can do without proper modelling of the quantum dissipative dynamics in presence of phase coexistence.  In this work we shall address right this issue in the prototype quantum first order phase transition displayed by a fully connected quantum Ising model with $p$-spin exchange, where $p>2$. \\
Quantum spin models,  besides being paradigmatic systems for studying quantum
phase transitions, also constitute a good playground  to investigate, both theoretically and  experimentally, 
the driven dissipative dynamics\cite{Wald_2016,PRATAVIERA2014,PhysRevLett.101.105701,PhysRevE.96.053306,PhysRevE.94.062143,PhysRevA.95.012122,Prosen_2009,PhysRevLett.119.150402,schwager}, including its realization in pertinently designed quantum impurity models realized at junctions of spin chains\cite{PhysRevB.72.014417,Giuliano_2008,GIULIANO2016135,MULLER2016482,PhysRevB.96.155145} .
We shall model the dissipative dynamics of our case study in the framework
of Markovian dynamics, through the rather general master equation
derived by Lindblad back in 1976 \citep{lindblad,pearle}, but  
still widely used \citep{nalbach,PhysRevE.95.042137,znidaricqubit,arceci,PhysRevB.74.144423,minganti,Teng_2006,PhysRevB.98.064307}.
As a matter of fact, metastability in Markovian open quantum systems turns out to be a non trivial problem \citep{Penrose1987-PENTAR} due to the separation of timescales in the dissipative dynamics \citep{PhysRevLett.116.240404}. Typically, the Lindblad equation is able to describe the short time dynamics during which the system relaxes to a metastable state \citep{PhysRevE.94.052132,PhysRevA.92.012128,MetaBCS}, while it fails to describe the long time ergodic dynamics that drives relaxation to the true equilibrium state.\\
Here, we show how to describe the full dynamics in terms of an effective Lindblad equation valid in both regimes. The main idea is to write the master equation as a sum of competing terms, one for each phase  within the metastable manifold. Within this description both supercooling and Mpemba effect spontaneously emerges during the quantum dissipative dynamics.\\

The paper is organized as follows. In Section~\ref{The model Hamiltonian} we introduce
the quantum Ising model we shall investigate and its equilibrium phase diagram. In Section~\ref{Lindblad equation} we briefly discuss the Lindblad master equation to describe the
Markovian relaxation dynamic, and consider the simple case of a single spin-1/2 in a static or dynamic 
magnetic field. In Section~\ref{Dynamics} we specialise the Lindblad equation 
in the fully connected quantum Ising model, and show how it can be exploited to recover 
the equilibrium phase diagram. In Section~\ref{choice 2} we present a variant of the Lindblad equation 
appropriate in case of phase coexistence, and apply it to our quantum Ising model, which indeed shows 
Mpemba effect. Finally, Sec.~\ref{Conclusions} is devoted to concluding remarks.

\section{The model Hamiltonian}
\label{The model Hamiltonian}

\begin{figure}
\centering{}\includegraphics[scale=0.2]{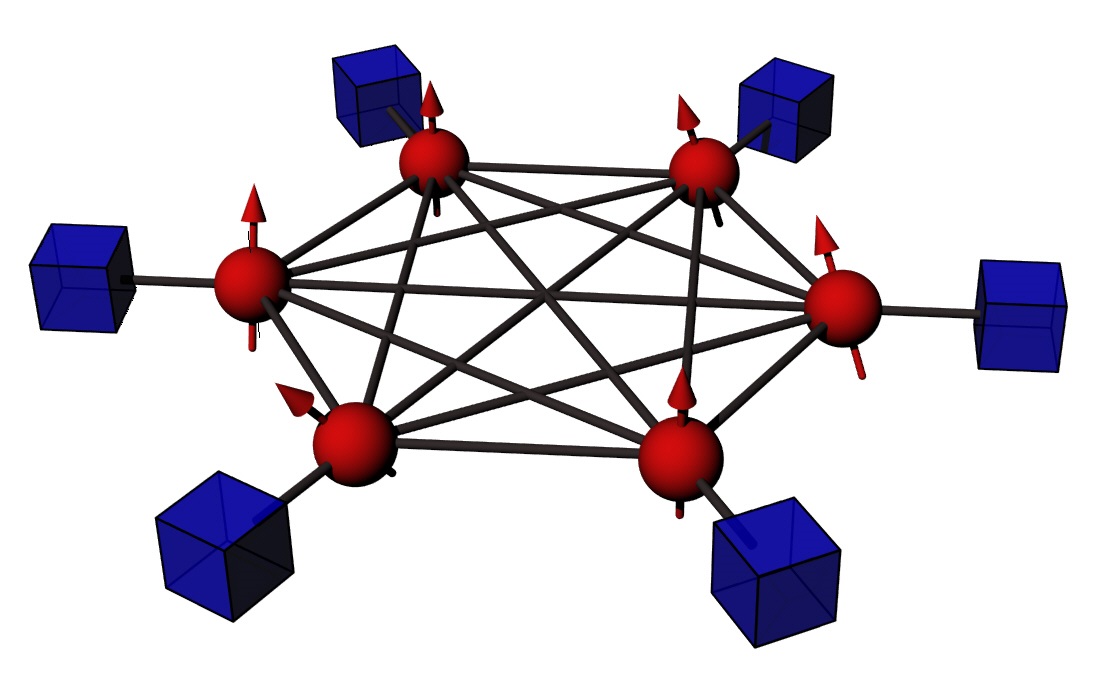}\caption{\label{fig:system} Graphic representation of the model: a fully connected graph of
$N$ spins (red balls), here shown for $N=6$, each in contact with its own bath (blue
boxes).}
\end{figure}

We wish to describe the dynamics of an open quantum
system, which must be as simple as possible but still posses a non-trivial phase diagram with metastable phases. For that purpose, we
consider a quantum Ising model on an $N$-sites
fully connected graph, sketched in Fig.~\ref{fig:system} for $N=6$,  
described by the Hamiltonian
\begin{eqnarray}
H & = & -h_{x}\sum_{i}\sigma_{i}^{x}-N\sum_{n=2}^{m}J_{n}\left(\frac{1}{N}\sum_{i}\sigma_{i}^{z}\right)^{n}\,,\label{Ham}
\end{eqnarray}
with integer $m\geq2$. Here $\sigma_{i}^{\alpha}$, $\alpha=x,y,z$, are Pauli matrices on site $i=1,\ldots,N$, $h_{x}$ a transverse magnetic
field, and $J_{n}$ $n$-spin exchange 
constants. 

\begin{figure}
\centering{}\includegraphics[width=0.45\textwidth]{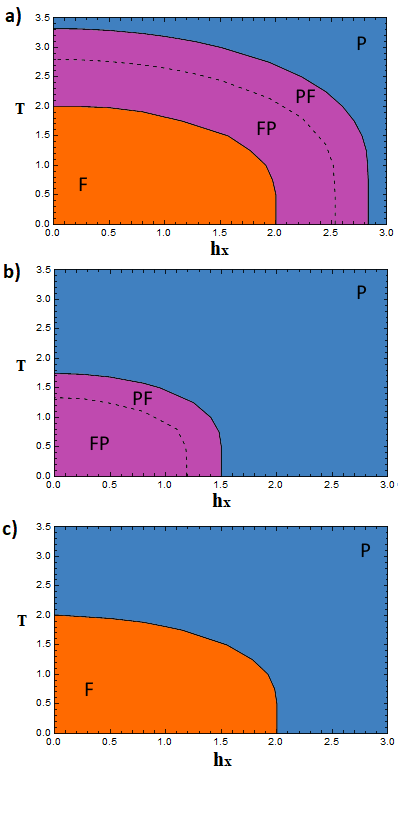}
\vspace{-1cm}
\caption{\label{ph-dia}
Phase diagram of the Hamiltonian \eqn{Ham} in case 1 with $J_2=J_4=1$, panel 
a), case 2 with $J_3=1$, panel b), and case 3 with $J_2=1$, panel c). 
The label F stand for the ferromagnetic phase with $m_z\not=0$, 
P for the paramagnetic one with $m_z=0$. FP denotes the coexistence region where the stable phase is F and the metastable P; the opposite case is denoted as PF. The first order transition between FP and PF in panels a) and b) is indicated as a dashed line, while the direct F to P transition in panel c) is 
a second order one. The F/FP and PF/P lines are, respectively, 
the P and F spinodal lines.    
}
\end{figure}

Hereafter, we shall focus on three reference
cases that are representative of all others. Specifically, 
\begin{enumerate}[label=\textbf{\arabic*.},ref=\arabic*]
\item $J_{2}\neq0$,
$J_{4}\neq0$, and $J_{n\neq2,4}=0$;\label{case1}
\item $J_{3}\neq0$ and $J_{n\neq3}=0$;\label{case2}
\item  $J_{2}\neq0$ and $J_{n\neq2}=0$.\label{case3}
\end{enumerate}
In all three cases the model undergoes a phase transition increasing either temperature $T$ or 
transverse field $h_x$ from a phase with finite to one with vanishing expectation value of  $\sigma^z_i$, $\forall~i$. This transition is first order in case 1 \cite{delre,Teng_2006} and 2 \cite{wauters}, but second order in case 3 \cite{PhysRevB.74.144423}. Moreover, in case 
1 and 3 it corresponds to the restoration of the $Z_2$ symmetry $\sigma^z_i\to -\sigma^z_i$, $\forall~i$, 
which is spontaneously broken in the low $T$ and $h_x$ phase, while such symmetry is explicitly broken in case 2. \\
Because of full connectivity, mean-field approximation becomes exact for model \eqn{Ham} 
in the thermodynamic limit $N\to\infty$, since, for 
$i\not=j$, and for any $\alpha,\beta=x,y,z$,   
\beal
\langle\, \sigma^\alpha_i\,\sigma^\beta_j\,\rangle - 
\langle \,\sigma^\alpha_i\,\rangle\,\langle\,\sigma^\beta_j\,\rangle \propto \fract{1}{N}\;
\underset{N\to\infty}{\longrightarrow}\,0\,.
\eal
It follows that the equilibrium Boltzmann distribution 
\beal
\rho = \fract{\esp{-\beta H}}{\Tr\Big(\esp{-\beta H}\Big)}\;
\underset{N\to\infty}{\longrightarrow}\;\prod_i\,\rho_i =
\prod_i\,\fract{\esp{-\beta H_i}}{\Tr\Big(\esp{-\beta H_i}\Big)}\;,\label{rho}
\eal
where $\beta=1/T$ and the local mean field Hamiltonian reads
\begin{eqnarray}
H_i & = & -h_{x}\,\sigma_{i}^{x} -h_z\big(\mb\big)\,\sigma_{i}^{z}\,.\label{Ham}
\end{eqnarray}
The longitudinal field  in \eqn{Ham} is defined by 
\begin{equation}
h_z\big(\mb\big)=\sum_{n=2}^{m}\,n\,J_{n}\, m_z^{n-1}\,,\label{eq:hzmean}
\end{equation}
where $\mb = \big(m_x,m_y,m_z\big)$ is the Bloch vector with components 
\begin{equation}
m_\alpha  \equiv\frac{1}{N}\sum_{j}\,
\left\langle \,\sigma_{j}^{\alpha}\,\right\rangle\,,\quad 
\alpha=x,y,z\,.
 \label{eq:meanspin}
\end{equation}
\begin{figure}
\centering{}\includegraphics[scale=0.45]{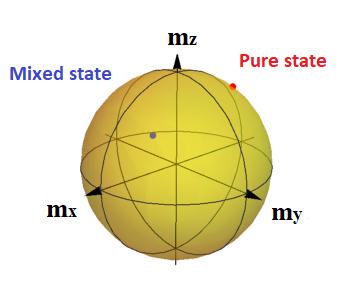}\caption{\label{fig:bloch sphere}Effective 2-d Bloch sphere for the mean values
of the spin components}
\end{figure}
The local density matrix $\rho_i$ in Eq.~\eqn{rho} can be also written as the $2\times2$ matrix
\begin{equation}
\rho_i=\frac{1}{2}\,\Big(\,1+\mb\cdot\bd{\sigma}_i\,\Big)
\,,\label{rho_i}
\end{equation}
where $\bd{\sigma}_i=\big(\sigma^x_i,\sigma^y_i,\sigma^z_i\big)$, and it is completely determined by the Bloch
vector $\mb$. 
In general $\left|\mb \right|\leq1$, being 
$\left|\mb \right|=1$ only 
for pure states (see Fig.~\ref{fig:bloch sphere}). Assuming the system at equilibrium with a bath at
at temperature $T$, the expectation values in \eqn{rho_i} read
\beal
\mb &= \tanh\big(\beta h(\mb)\big)\,\Big(\cos\theta(\mb)\,,\,0\,,\,\sin\theta(\mb)\Big)\,,
\label{eq:equilibrium}
\eal
where 
\beal
\tan\theta\big(\mb\big) &= \fract{\;h_z(\mb)\;}{h_x}\;,\\
h(\mb)&= 
\sqrt{\,h_x^2 + h_z(\mb)^2\;}\;.\label{self-con}
\eal
Eqs.~\eqn{eq:equilibrium} and \eqn{self-con} self-consistently determine the equilibrium state at 
temperature $T$, leading to the phase diagram shown in Fig.~\ref{ph-dia} for the three different cases. In the figure, F denotes the ferromagnetic phase with $m_z\not=0$, P the paramagnetic one with $m_z=0$, 
while FP and PF the coexistence regions, FP with F stable and P metastable and PF vice versa.  

\section{Lindblad equation}
\label{Lindblad equation}

The dynamics of open quantum systems is usually described through 
a master equation, aimed to represent the true quantum evolution after integrating out the bath degrees of freedom. The derivation of such equation usually relies on the so-called Markovian approximation, which consists of neglecting memory effects under the assumption that the bath relaxation-time is much shorter
than the characteristic time-scales of the system.
The Lindblad equation is among the most used master equations\cite{lindblad}. 
It can be derived in various ways and under different assumptions, like, e.g., the quantum dynamical
semigroups formalism, the Ito stochastic calculus \cite{ADLER200058}, the projector techniques
associated to the rotating wave approximation \cite{REDFIELD19651} or the Keldysh diagrammatic
formalism\cite{PhysRevA.95.013847,Sieberer_2016}. \\
If the Born approximation is valid, i.e., if we can safely neglect system-bath
correlations, the system (S) + bath (B) density matrix can be written as a tensor product 
\begin{equation}
\rho_{S+B}\left(t\right)\simeq \rho_{S}\left(t\right)\,\otimes\,\rho_{B}\left(t\right)\,. 
\end{equation}
The general form of the Lindblad equation consists in a first-order differential
equation for the time evolution of the system density matrix $\rho_S(t)$:
\beal
\dot{\rho}(t)_{S}& =-i\,\Big[H_{S}\,,\,\rho_{S}(t)\Big]\\
&\quad +\sum_{\lambda}\,\Bigg[
\gamma_{\lambda}\,\bigg(2L_{\lambda}\,\rho_{S}\,L_{\lambda}^{\dagger}
-\Big\{ L_{\lambda}^{\dagger}\,L_{\lambda}\,,\,\rho_{S}\Big\} \bigg)\Bigg]\,, \label{eq:lindblad}
\eal
where $H_{S}$ is the system Hamiltonian and $L_{\lambda}$ are the 
so-called Lindblad or jump operators, which are determined by the coupling between the system and the bath,  
and span the space of all independent operators within the system Hilbert
space. The first term in the Lindblad equation \eqn{eq:lindblad} is the so-called Liouvillian
that describes the unitary evolution brought by $H_S$, while the
second term, so-called Lindbladian, includes dissipation and decoherence in the dynamics. Neglecting pure dephasing processes, the relaxation dynamics of a system is described by non-hermitian
jump operators that produce transitions between the 
eigenstates $\ket{n}$ of $H_{S}$ with eigenvalues $E_n$. Specifically, 
we shall define\citep{pearle}   
\beal
L_{\lambda\left(m,n\right)} &=\left|m\right\rangle \left\langle n\right|\,,
& E_n &< E_m \,.\label{def:L}
\eal
Using such definition of $L_\lambda$, we can write Eq.~\eqn{eq:lindblad} as 
\beal
\dot{\rho}_{S}&=-i\Big[H_{S}\,,\,\rho_{S}\Big]+\sum_{\lambda}\,\Bigg[ \\
&\qquad \gamma_{\lambda}\bigg(2L_{\lambda}\,\rho_{S}\,L_{\lambda}^{\dagger}-\Big\{ L_{\lambda}^{\dagger}\,L_{\lambda}\,,\,\rho_{S}\Big\} \bigg)\\
&\qquad +\bar{\gamma}_{\lambda}\bigg(2L_{\lambda}^{\dagger}\,\rho_{S}\,L_{\lambda}
-\Big\{ L_{\lambda}L_{\lambda}^{\dagger}\,,\,\rho_{S}\Big\} \bigg)\Bigg]\,. \label{eq:lindblad-rel}
\eal
One can readily verify that the Boltzmann distribution is a stationary solution of \eqn{eq:lindblad-rel} 
if the coupling constants $\gamma_{\lambda}$ and 
$\bar \gamma_{\lambda}$ are related to each other through 
\begin{equation}
\fract{\bar \gamma_{\lambda}}{\gamma_{\lambda}}=\esp{\beta\,\epsilon_{\lambda}}\;,
\label{cond-gamma}
\end{equation}
where $\epsilon_{\lambda}=E_{m}-E_{n}>0$ are excitation energies. 
In what follows, we shall use the Lindblad equation \eqn{eq:lindblad-rel} with the condition \eqn{cond-gamma}.

\subsection{A single spin-1/2 in a magnetic field}
\label{A single spin-1/2 in a magnetic field}

As an example, useful in the next discussion of the Hamiltonian \eqn{Ham}, let us consider the paradigmatic case of a single spin-1/2 in a magnetic field,  
with Hamiltonian
\begin{equation}
H_{S}=-\mathbf{h}\cdot\bd{\sigma}
= -\big|\bd{h}\big|\;\bd{v}_3\cdot\bsigma
\,,\label{eq:hx}
\end{equation}
where $\bd{v}_3\,||\,\bd{h}$ is a unit vector,
coupled to a dissipative bath at temperature $T$. In this simple case, the ground state of $H_S$ satisfies the eigenvalue equation $H_S\ket{0}=-|\mathbf{h}|\ket{0}$, and there is a single excited state, $\ket{1}$, at energy $\ep=2|\mathbf{h}|$ above. 
It follows that there is only one jump operator defined through  Eq.~\eqn{def:L},  
namely
\beal
L & = \, \ket{1}\bra{0} \,= \big(\bd{v}_1 - i\,\bd{v}_2\big)\cdot\bd{\sigma}/2\equiv \bd{v}^-\cdot\bd{\sigma}/2\,,\label{L:sigma^-}
\eal
with $\epsilon_\lambda=2|\bd{h}|$, where $\bd{v}_1$ and $\bd{v}_2$ are real orthogonal unit vectors satisfying $\bd{v}_1\wedge\bd{v}_2\cdot\bd{v}_3 = 1$. Plugging $L$, and its hermitian conjugate, 
\beal
L^\dagger &= \, \ket{0}\bra{1} \,=\big(\bd{v}_1 + i\,\bd{v}_2\big)\cdot\bd{\sigma}/2\equiv 
\bd{v}^+\cdot\bd{\sigma}/2\,,\label{L:sigma^+}
\eal
into 
Eq.~\eqn{eq:lindblad-rel} and computing the expectation values of the spin
operator we obtain the Lindblad equation for the magnetisation $\mb(t)$, namely, 
\beal
&\dot{\bd{m}}(t) \equiv \Tr\Big(\dot\rho_S(t)\,\bd{\sigma}\Big)
= -2\bd{h}\wedge\bd{m}(t) \\
& \;-\fract{\gamma}{2}\,\bigg[4\Big(\bd{v}_3+\mb(t)\Big)
- \bd{v}^-\,\Big(\bd{v^+}\cdot\mb(t)\Big) \\
&\qquad\qquad\qquad\qquad\qquad 
- \bd{v}^+\,\big(\bd{v^-}\cdot\mb(t)\big)\bigg]\\
&\; +\fract{\bar \gamma}{2}\,\bigg[4\Big(\bd{v}_3-\mb(t)\Big)
+ \bd{v}^-\,\Big(\bd{v^+}\cdot\mb(t)\Big)\\
&\qquad\qquad\qquad\qquad\qquad  
+ \bd{v}^+\,\Big(\bd{v^-}\cdot\mb(t)\Big)\bigg]\,.
\label{eq:ising-spin-equ}
\eal
The stationary solution correctly reproduces the thermal equilibrium, 
$\mb\cdot\bd{v}_3 = \tanh(\beta\,|\bd{h}|)$, and $\mb\cdot\bd{v}_1
=\mb\cdot\bd{v}_2=0$. Eq.~\eqn{eq:ising-spin-equ} 
allows following the system evolution within the Bloch sphere
from an arbitrary initial condition to the equilibrium stationary state. 

Let us consider now the same system Hamiltonian but now in presence 
of a time dependent magnetic field 
\begin{equation}
H_{S}\big(\bd{h}(t)\big)=-\bd{h}(t)\cdot\bsigma 
= -\big|\bd{h}(t)\big|\,\bd{v}_3(t)\cdot\bsigma
\,,\label{Ham-1}
\end{equation}
where $\bd{h}(t)$ evolves from an initial value, $\bd{h}(t\leq 0)=\bd{h}_i$, to a final one, $\bd{h}(t\gg 1)=\bd{h}_f$.
A possible choice of the Lindblad operator that guarantees relaxation 
to the equilibrium density matrix $\rho_{S f}$ of the final Hamiltonian $H_S\big(\bd{h}_f\big)$
is the one in Eq.~\eqn{def:L} defined through the 
eigenstates of the instantaneous Hamiltonian, i.e., from Eq.~\eqn{L:sigma^-},
\beal
L(t)  &= \bd{v}^-(t)\cdot\fract{\bd{\sigma}}{2}\,,&
L^\dagger(t)  &= \bd{v}^+(t)\cdot\fract{\bd{\sigma}}{2}\,,
\label{L:sigma^-(t)}
\eal
where $\bd{v}^+(t)=\bd{v}^-(t)^*$, which satisfy 
\be
\bd{v}^+(t)\wedge\bd{v}^-(t)=2\bd{v}_3(t)\,.
\label{cond-1}
\ee
The Lindblad equation is the same as in Eq.~\eqn{eq:ising-spin-equ}, 
though with time dependent $\bd{v}^{\pm}(t)$ and $\ep_\lambda(t)=2\big|\bd{h}(t)\big|$ in \eqn{cond-gamma}.\\
In reality, the precise time dependence of the Lindblad operator 
$L(t)$ is not crucial to guarantee relaxation to the equilibrium $\rho_{Sf}$; 
what actually matters is just that $L(t)$ 
becomes, for $t\gg 1$, the Lindblad operator \eqn{L:sigma^-}
corresponding to the final Hamiltonian $H_s(\bd{h}_f)$. 

\section{Dynamics in the fully connected quantum Ising model}
\label{Dynamics}

Let us consider a generic model in contact with a bath and subject to a quench of the state variables or of the Hamiltonian parameters, such that its phase diagram with the final system parameters comprises metastable phases besides the stable one.  
In this case, the proper choice of the Lindblad operators that 
could describe equally well the approach to equilibrium and  
non equilibrium phenomena like, e.g., supercooling or superheating, is 
not so straightforward. For instance, the most natural choice of Lindblad operators defined as in Eq.~\eqn{def:L} through the eigenstates of 
the final Hamiltonian at the final values of the state variables does yield relaxation to equilibrium, 
but cannot produce trapping in a metastable phase.\\

We shall tackle precisely this issue in the 
simple model Hamiltonian \eqn{Ham}.
Here, full connectivity ensures that also 
the time-dependent density matrix of the system becomes factorizable 
in the thermodynamic limit, $N\to\infty$, namely,  
\be
\rho_S(t) \;\underset{N\to\infty}{\longrightarrow}\;
\prod_i\,\rho_i(t)\,,
\ee
where $\rho_i(t)$ describes the time evolution of the spin at site $i$ 
coupled to a bath at temperature $T$ and in presence of an effective time dependent magnetic field, exactly like the Hamiltonian \eqn{Ham-1}. The difference with the latter is that, through \eqn{eq:hzmean}, the field 
\beal
\bd{h}(t) =\bd{h}\big(\mb(t)\big) &= \bigg(h_x\,,\,0\,,\,\sum_{n=2}^m\,n\,J_n\,m_z(t)^{n-1}
\bigg)\\
&\equiv \Big|\bd{h}\big(\mb(t)\big)\Big|\;\bd{v}_3\big(\mb(t)\big)\,,\label{eff-h(t)}
\eal
with 
\be
\mb(t) = \fract{1}{N}\,\sum_i\,\Tr\Big(\rho_S(t)\,\bsigma_i\Big)\,,
\ee
is self-consistently determined by the system time evolution, and thus the 
Lindblad equation acquires non-linear terms. \\
In this case we have several options for choosing the Lindblad jump operators. For instance, we may define $L(t)$ by the eigenstates of the instantaneous system Hamiltonian, namely using the expression in Eq.~\eqn{L:sigma^-(t)}, where 
$\bd{v}_3(t)$ in Eq.~\eqn{cond-1} is defined by Eq.~\eqn{eff-h(t)} and 
$\ep_\lambda=2\big|\bd{h}\big(\mb(t)\big)\big|$. \\
This is however not the only choice, as there are several alternative definitions of the jump operators.  
We already showed in Sec.~\ref{The model Hamiltonian} that the 
free energy of the model Hamiltonian \eqn{Ham} 
has different minima, characterised by different values $m_z^{(a)}$, $a=1,\dots$, of $m_z$, and thus each of them associated with a 
different effective field 
\be
\bd{h}^{(a)} = \bigg(h_x\,,\,0\,,\,\sum_{n=2}^m\,n\,J_n\,m^{(a)}_z{^{n-1}}
\bigg)\equiv \big|\bd{h}^{(a)}\big|\,\bd{v}_3^{(a)}\,.\label{h-a}
\ee
We can thus choose as jump operators those 
associated with the free energy minima $a=1,\dots$, namely, 
\beal
L^{(a)} &= \bd{v}{^{(a)}}{^-}\cdot\fract{\bd{\sigma}}{2}\;,& 
L^{(a)}{^\dagger} &= \bd{v}{^{(a)}}{^+}\cdot\fract{\bd{\sigma}}{2}\;,
\label{L-a}
\eal 
where 
\be
\bd{v}^{(a)}{^+}\wedge\bd{v}{^{(a)}}{^-}\equiv 
\big(\bd{v}{^{(a)}}{^-}\big)^*\wedge\bd{v}{^{(a)}}{^-}=2\bd{v}_3^{(a)}\,,
\label{cond-2}
\ee
and Eq.~\eqn{cond-gamma} with 
\be
\ep_\lambda=\ep_a =2\Big|\bd{h}^{(a)}\Big|\,.
\ee

In what follows, we analyse strengths and weaknesses of $L(t)$ defined through the eigenstates of the instantaneous Hamiltonian, while in the next section we shall present an alternative definition based on the jump operators \eqn{L-a} that yields physically more sound dynamics in presence of metastable phases.

\subsection{Jump operator defined through the instantaneous Hamiltonian}
\label{choice 1}
We study first the dynamical evolution with the jump operator 
$L(t)$ defined through the equations \eqn{L:sigma^-(t)}, \eqn{eff-h(t)} and 
\eqn{cond-1}. The Lindblad equation for the average magnetisation $\mb(t)$ 
is the same as in Eq.~\eqn{eq:ising-spin-equ}, with time dependent 
$\bd{h}\big(\mb(t)\big)$ and $\bd{v}^{\pm}\big(\mb(t)\big)$. One readily realises that the stationary 
solutions of that equation, including or not the Liouvillian, i.e., the first term on the r.h.s. of Eq.~\eqn{eq:ising-spin-equ}, correspond to the extrema of the free energy. The 
magnetisation $\mb(t)$ thus flows towards one of those extrema depending on the initial conditions. More rigorously, $\mb(t)$ flows towards one of the  minima, unless it is initially right on a saddle point. We can therefore integrate numerically Eq.~\eqn{eq:ising-spin-equ} starting from any initial condition and, by doing so, map out all basins of attraction within the Bloch sphere, 
which is actually the same as calculating the phase diagram.

\begin{figure}
\centering{}\includegraphics[scale=0.10]{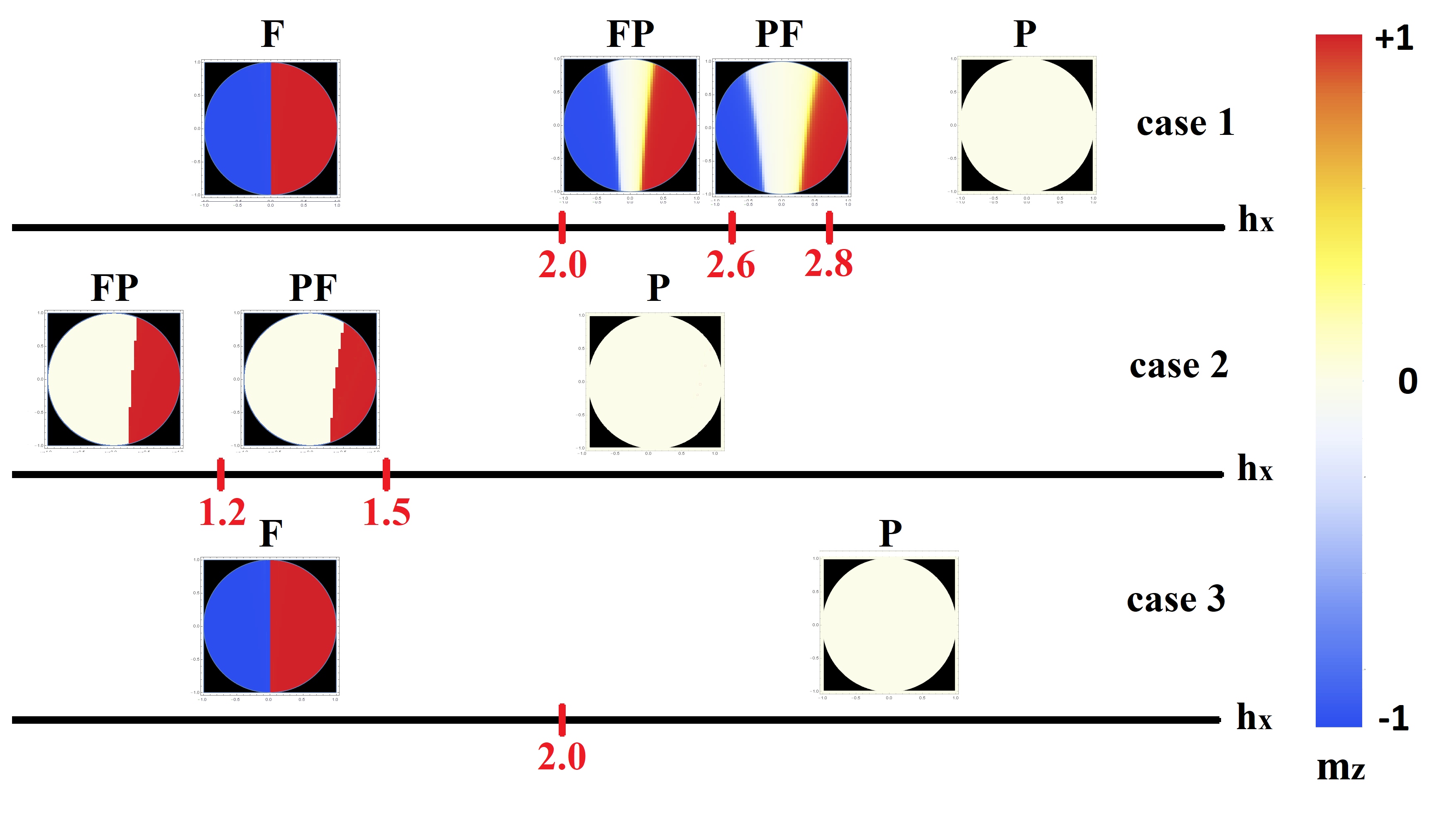}\caption{\label{fig:diagramnoliouvillian}
Basins of attraction of the Lindblad equation neglecting the Liouvillian for the magnetisation $\mb(t)$  within the Bloch sphere projected onto the 
$m_y=0$ plane and for $T\to 0$.  
Top panel: case 1 with $J_2=J_4=1$. Middle panel: case 2 
with $J_3=1$. Bottom panel: case 3 with $J_2=1$. The colours indicate the stationary value of the longitudinal magnetisation $m_z$; specifically, red 
means $m_z>0$, blue $m_z<0$ and white $m_z=0$.}
\end{figure}
\begin{figure}
\centering{}\includegraphics[scale=0.10]{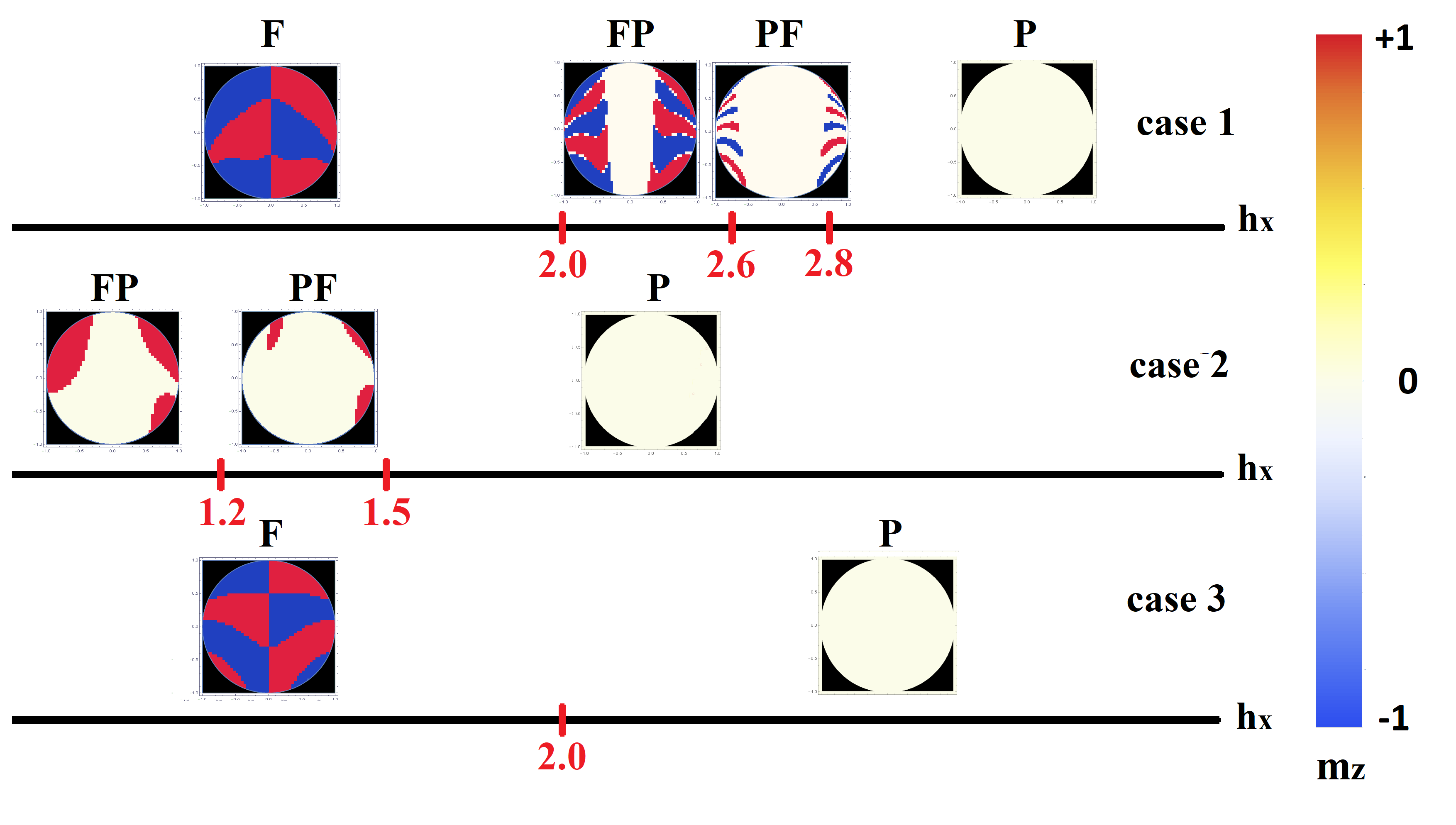}\caption{\label{fig:diagramsiliouvillian}
Same as Fig.~\ref{fig:diagramnoliouvillian} including the Liouvillian. Note 
that apparently disconnected regions with the same colour are actually connected along the $y$-direction.}
\end{figure}
In Figs.~\ref{fig:diagramnoliouvillian} and \ref{fig:diagramsiliouvillian} we show the basins of attraction 
of Eq.~\eqn{eq:ising-spin-equ} discarding and including, respectively, the 
Liouvillian, at very small $T$ and in the three cases of interest. 
Considering for instance case 1, we note at small $h_x$ two basins of attraction at finite and opposite values of $m_z$, which signal the 
$Z_2$ symmetry broken phase. The three basins for intermediate values of $h_x$ indicate instead the coexistence regions, FP or PF. Finally, at large $h_x$ 
there is only one basin with $m_z=0$.   
Repeating the above analysis at any temperature, 
we can obtain the same phase diagrams of Fig.~\ref{ph-dia}. 

We end noting that, while the Lindblad jump operator defined through the eigenstates of the instantaneous Hamiltonian provides an alternative and efficient way to calculate the phase diagram, it fails 
to produce relaxation to equilibrium. Indeed, when the system starts in the basin of attraction of a metastable phase, it will remain trapped there forever, at least in our simple model \eqn{Ham}.   

\section{Lindblad operators in presence of phase coexistence}
\label{choice 2}
The unsatisfying results in Sec.~\ref{choice 1} call for a different 
definition of jump operators able to describe the expected relaxation to equilibrium even in presence of metastable phases. For that, we observe 
that the Hilbert space of a system that undergoes phase transitions comprises
several disconnected subspaces. The thermodynamic properties at given temperature and Hamiltonian parameters is determined just by one of those 
subspaces at a time. Let us consider, as an example, a system whose phase diagram looks like panel c) in Fig.~\ref{ph-dia}, which displays a second order transition into a broken symmetry phase. 
In this case the Hilbert space in the thermodynamic limit decomposes into two subspaces; one that includes the symmetry invariant eigenstates, and the other the symmetry variant ones. The latter is in turn further decomposed into sub subspaces, also disconnected in the thermodynamic limit, and related one to the other by the generators of the symmetry that is spontaneously broken.
A similar decomposition holds also when the transition is first order, 
like in cases 1 and 2 of the Hamiltonian \eqn{Ham}, whose phase diagrams 
are shown in panels a) and b), respectively, of Fig.~\ref{ph-dia}. 
The difference here is that the disconnected subspaces overlap in energy, thus the coexistence regions. 

In all these situations it is rather natural to introduce jump operators of the form in Eq.~\eqn{def:L} within each subspace, since different subspaces  are disconnected from each other. In a dissipative environment, each subspace will act as a basin of attraction of the dynamics, which implies that the Lindblad equation would generally include jump operators of all subspaces, namely
\beal
\dot{\rho}_{S}&=-i\Big[H_{S}\,,\,\rho_{S}\Big]+
\sum_a\,\alpha_a(t)\sum_{\lambda}\,\Bigg[ \\
&\quad \gamma^{(a)}_{\lambda}\bigg(2L^{(a)}_{\lambda}\,\rho_{S}\,L^{(a)}_{\lambda}{^{\dagger}}-\Big\{ L^{(a)}_{\lambda}{^{\dagger}}\,L^{(a)}_{\lambda}\,,\,\rho_{S}\Big\} \bigg)\\
&\; +\bar{\gamma}^{(a)}_{\lambda}\bigg(2L^{(a)}_{\lambda}{^{\dagger}}\,\rho_{S}\,L^{(a)}_{\lambda}
-\Big\{ L^{(a)}_{\lambda}\,L^{(a)}_{\lambda}{^{\dagger}}\,,\,\rho_{S}\Big\} \bigg)\Bigg]\,. \label{def-L-a-b}
\eal
where $a=1,\dots$ labels the subspaces, and the coefficient $\alpha_a(t)$ weighs the attraction strength of subspace $a$.
Physically, we expect that $\alpha_a(t)$ is (1) smaller for 
subspaces that correspond to metastable phases than for those corresponding to stable ones; and (2) smaller   
the closer the system instantaneously is to the basin of attraction of another subspace. These two features guarantee that the system does flow to the 
equilibrium stable phase, unless it happens to be deep in the basin of attraction of a metastable phase, and thus remedy the drawbacks of 
the infinite lifetimes of metastable phases highlighted 
in Sec.~\ref{choice 1}, without spoiling supercooling or superheating. 
Incidentally, we note that choosing $\alpha_a(t)=1$ and $\alpha_{b\not=a}=0$
in Eq.~\eqn{def-L-a-b} brings to a steady state trapped into the subspace 
$a$, even if it not the equilibrium one.\\  

In the simple case of the mean-field Hamiltonian \eqn{Ham}, the 
different subspaces actually corresponds to the different minima of the 
free energy that we have discussed in Sec.~\ref{Dynamics}. Moreover, 
the above requirements (1) and (2) can be easily implemented. For instance, 
we may assume for the coefficients $\alpha_a(t)$ in Eq.~\eqn{def-L-a-b} 
the expression   
\beal
\alpha_a(t) &= \Gamma\,\esp{-\beta f_a}\;\prod_{b\not=a}\,\Big(\,
1-\bd{v}_3\big(\mb(t)\big)\cdot\bd{v}_3^{(b)}\,\Big).\label{alpha-a}
\eal
Here $\Gamma$ determines the overall strength of the coupling with the bath,    
while $\bd{v}_3\big(\mb(t)\big)$ and $\bd{v}_3^{(a)}$ are defined in equations 
\eqn{eff-h(t)} and \eqn{h-a}, respectively, and implement the condition (1). Finally, $\esp{-\beta f_a}$ is an Arrhenius term that involves the free energy per site $f_a$ of the minimum $a$, and enforces the condition (2). 

\subsection{Mpemba effect}
The Lindblad equation \eqn{def-L-a-b}, with $\alpha_a(t)$ defined in 
Eq.~\eqn{alpha-a}, yields a physically sound dissipative dynamics of the 
Hamiltonian \eqn{Ham}, including also the Mpbemba effect discussed in the Introduction.  
\begin{figure}
\centering{}\includegraphics[scale=0.33]{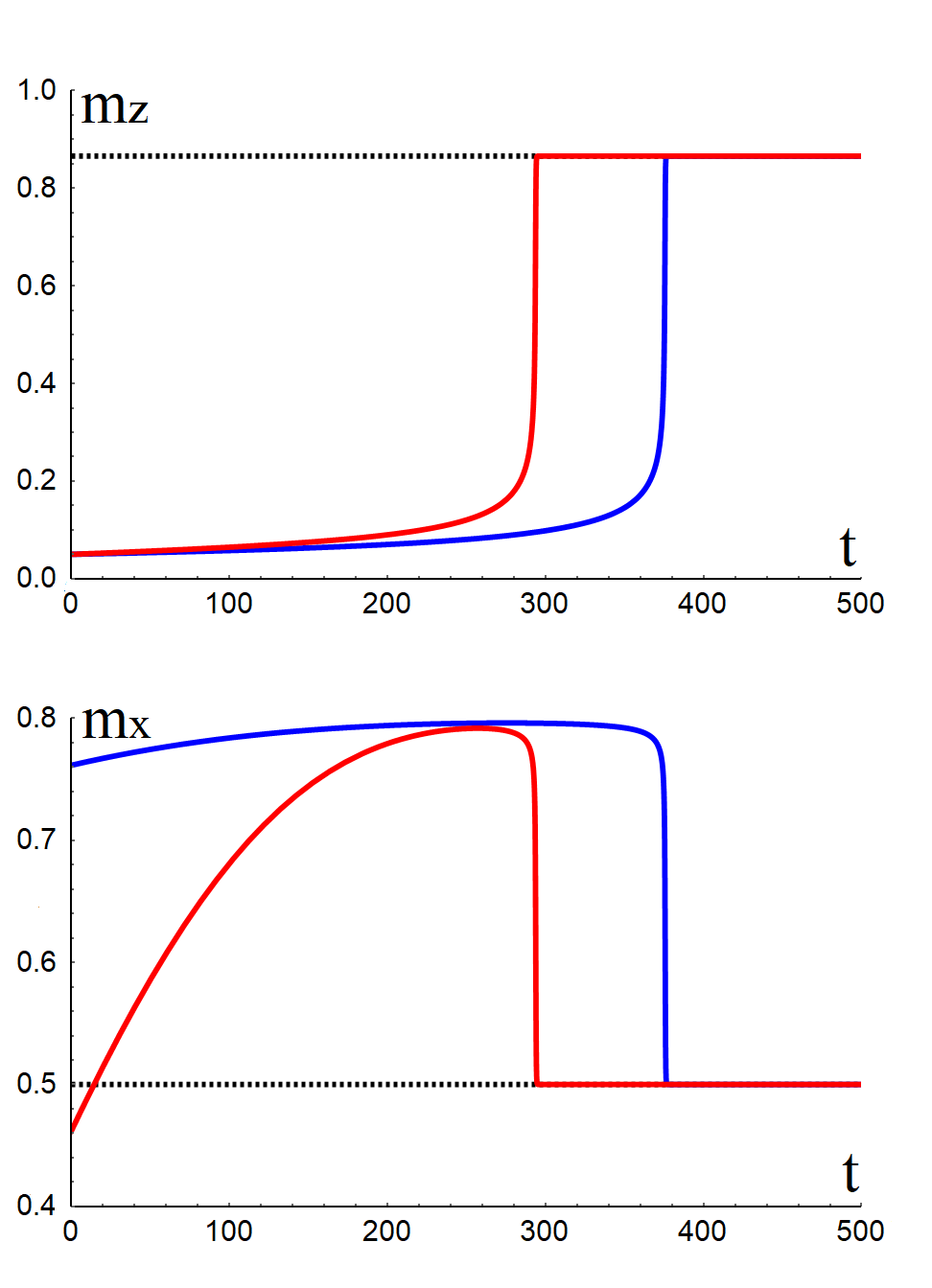}\caption{\label{mpemba} Dissipative dynamics of 
the Hamiltonian \eqn{Ham} in case 1 with $J_2=J_4=1$ and $h_x=2.5$, 
see phase diagram in panel a) Fig.~\ref{ph-dia}, after a sudden quench of 
temperature from $T_i$ to $T_f=0.002\ll T_i$. Blue and red lines refer 
to $T_i=2.5$ and $T_i=5$, respectively. Top panel: longitudinal magnetisation 
$m_z$. Bottom panel: transverse magnetisation $m_x$.  
We observe that the initially hotter system (red lines) reaches the equilibrium value of magnetisation faster than the colder one (blue lines).}
\end{figure}
For that, let us consider the Hamiltonian \eqn{Ham} in case 1 with 
$J_2=J_4=1$ and $h_x=2.5$, see panel a) in Fig.~\ref{ph-dia}. 
We assume two copies, one initially at equilibrium with a bath at temperature
$T_{i,1}$, and the other at temperature $T_{i,2}$, with $T_{i,1}<T_{i,2}$ 
and both above the F spinodal point $T_{sF}\simeq 2.4$. Therefore, at first both copies are in the P phase, panel a) in Fig.~\ref{ph-dia}. 
We then quench both systems to the same final temperature $T_f$ 
that falls into the FP coexistence region, with the ferromagnetic phase 
stable and the paramagnetic one metastable. In order to seed nucleation of the F phase within the P one, we assume initially a tiny but finite $m_z=0.05$.  

In Fig.~\ref{mpemba} we show the time evolution obtained by integrating the 
Lindblad equation of the longitudinal magnetisation $m_z$, top panel, 
and transverse magnetisation $m_x$, bottom panel, both for the initially 
hotter (red lines) and colder (blue lines) copies. We observe a quite long transient where both copies remain trapped into the P phase they started from, metastable at the final temperature, which can thus be viewed as 
a supercooled P phase. However, the hotter copy gets out of the metastable P 
phase, and fast reaches thermalisation, earlier than the colder one; this is just the Mpemba effect.

\section{Conclusions}
\label{Conclusions}
In this work we have studied several variants of the Lindblad equation to 
describe the dissipative dynamics in presence of phase coexistence. 
In particular we have considered the exactly-solvable fully-connected 
quantum Ising model with two and four spin-exchange, whose phase diagram is 
the prototype of a symmetry-breaking quantum first order transition. 
We have found that a sound dissipative dynamics is recovered only when    
the Lindblad equation involves jump operators defined in each of the coexisting phases, whether stable or metastable. Applying such equation to the fully-connected quantum Ising model, we are able to describe also 
intriguing non-equilibrium phenomena, like the Mpemba effect.  

\section*{ACKNOWLEDGMENTS}

We acknowledge discussions with M. {\v{Z}}nidari{\v{c}}, G. Santoro, T. J. G.
Apollaro and L. Arceci. This work was supported by the European Union,
under ERC FIRSTORM, contract N. 692670.

\bibliography{lindblad_biblio}

\end{document}